\UseRawInputEncoding

\documentclass[10pt, prb, aps, twocolumn, showpacs, citeautoscript, floatfix, reprint, amsmath, amssymb, notitlepage,  superscriptaddress]{revtex4-1}

\usepackage{hyperref}
\usepackage{graphicx}
\usepackage{stmaryrd}
\usepackage{rotating}
\usepackage{amsmath}
\usepackage{amsfonts}
\usepackage{amssymb}
\usepackage[countmax]{subfloat}
\usepackage{dcolumn} 
\usepackage{bm} 
\usepackage{color}

\usepackage{float}
\usepackage{latexsym}

\usepackage{soul} 

\usepackage{macros}
\usepackage{braket}

\begin{document}

\title{Measurement of Topological Order based on Metric-Curvature Correspondence}




\author{Gero von Gersdorff}
\affiliation{Department of Physics, PUC-Rio, 22451-900 Rio de Janeiro, Brazil}


\author{Wei Chen}
\affiliation{Department of Physics, PUC-Rio, 22451-900 Rio de Janeiro, Brazil}

\date{\today}

\begin{abstract}

A unified expression for topological invariants has been proposed recently to describe the topological order in Dirac models belonging to any dimension and symmetry class. We uncover a correspondence between the curvature function that integrates to this unified topological invariant and the quantum metric that measures the distance between properly defined many-body Bloch states in momentum space. Based on this metric-curvature correspondence, a time-resolved and angle-resolved photoemission spectroscopy experiment is proposed to measure the violation of spectral sum rule caused by a pulse electric field to detect the quantum metric, from which the topological properties of the system may be extracted.


\end{abstract}

\maketitle

\section{Introduction}  

Topological order in materials differs from the usual Landau order parameters in that it does not require breaking a continuous symmetry, but rather represents a certain geometric property of the Bloch state in momentum space. In addition, topological order manifests as different physical phenomena according to the dimension and symmetry of the system\cite{Schnyder08,Ryu10,Kitaev09,Chiu16}, such as quantized Hall conductance\cite{Thouless82,Niu85} and Majorana fermions\cite{Kitaev01,Oreg10,Lutchyn10}, many of which rely on the metallic edge state that only exists at the boundary and in topologically nontrivial phases. Nevertheless, drawing analogy with Landau order parameters, an intriguing question is whether a bulk spectroscopy method exists to measure the geometric property of the Bloch state, through which the topological order can be detected. Should such a bulk spectroscopy exist ubiquitously for any topological insulator (TI) and topological superconductor (TSC), the aforementioned features seem to suggest that the detection principle cannot rely on the existence of local order parameter or edge state, which rules out many existing methods. 


Two recent progresses shed a light on this issue. The first is the recognition that for topological materials described by two-band Dirac models, the modulus of the Berry connection or Berry curvature that integrates to the topological invariant is equal to the quantum metric\cite{Provost80,Berry89} that measures the distance between single-particle Bloch state in momentum space\cite{Ma13,Kolodrubetz13,Ma14,Yang15,Piechon16,Kolodrubetz17,Ozawa18,Palumbo18,
Palumbo18_2,Lapa19,Yu19,Chen20_Palumbo,Ma20,Salerno20,Lin21}. As a result, direct measurement to the quantum metric may yield information about the topological order. The second is that all the topological invariants for Dirac models in any dimension and symmetry class\cite{Schnyder08,Ryu10,Kitaev09,Chiu16} can be unified into a single formula called wrapping number, which counts how many times the Brillouin zone (BZ) wraps around a target sphere induced by the Dirac Hamiltonian\cite{vonGersdorff21_unification}. In particular, the wrapping number is calculated from integrating the cyclic derivative of the components of the Dirac Hamiltonian, which we refer to as the curvature function. In two-band systems, the curvature function is simply the Berry connection or Berry curvature.

The goal of this Letter is to demonstrate that this equivalence between the modulus of the Berry connection or Berry curvature and the quantum metric in fact holds for Dirac models in any dimension and symmetry class. We elaborate that the modulus of the curvature function that integrates to the wrapping number corresponds to the quantum metric of a properly defined many-body Bloch state, a relation that we call metric-curvature correspondence. Motivated by this correspondence, we generalized a previously proposed measurement protocol for the single-particle quantum metric\cite{Ozawa18} to degenerate bands, and propose a  time-resolved and angle-resolved photoemission spectroscopy (trARPES) measurement\cite{Hajlaoui13,Sobota14,Lv19,Sobota14_2,Wang12_3} that detects the violation of spectral sum rule  caused by a pulse electric field as a universal spectroscopy to probe the topological property of materials, using graphene as a concrete example.

\section{Metric-curvature correspondence} 

\subsection{Theoretical formalism}

We consider the  TIs and TSCs described by the Hamiltonian and Bloch eigenstates $H({\bf k})|\psi_{n}({\bf k})\rangle=\eps_{n}({\bf k})|\psi_{n}({\bf k})\rangle$,  and focus on the filled valence band or  fermionic quasiparticle states with $\eps_{n}({\bf k})<0$. The $D$ dimensional BZ $T^D$ is parametrized by Cartesian coordinates $k^\mu$ with $\mu=1,2...D$, where the Einstein notation is used throughout the article. For a generic single or many particle Bloch state
$\ket {\psi(\vec k)}$
the overlap of this eigenstate at ${\bf k}$ with itself at a slightly different momentum $|\langle\psi({\bf k})|\psi({\bf k}+\delta{\bf k})\rangle|= 1-\frac{1}{2}g_{\mu\nu}\delta k^{\mu}\delta k^{\nu}$ defines the quantum metric tensor of that state  \cite{Provost80}
\begin{eqnarray}
g^{ \psi}_{\mu\nu}(\vec k)&=&\frac{1}{2}\langle\partial_{\mu}\psi|\partial_{\nu}\psi\rangle
+\frac{1}{2}\langle\partial_{\nu}\psi|\partial_{\mu}\psi\rangle\nn\\
&&-\langle\partial_{\mu}\psi|\psi\rangle\langle\psi|\partial_{\nu}\psi\rangle.
\label{gab_multiple_lambda}
\end{eqnarray} 
which is invariant under a local $U(1)$ gauge rotation $\ket{\psi(\vec k)}\to e^{i\phi(\vec k)}\ket{\psi(\vec k)}$. We remark that this metric tensor on the BZ is the one inherited from the well known Fubini-Study metric on state space \footnote{
For instance, in the case of one-particle Bloch states, this state space is $\mathbb{CP}^{N-1}$.
 Technically, $g_{\mu\nu}^\psi$ 
is the pullback of the Fubini-Study metric over the map $\ket{\psi(\vec k)}$.
}.


For reasons that will become transparent in a moment, we will consider the quantum metric constructed from the following $N_-$-particle Bloch state
\be
\ket{\psi^{\rm val}(\vec k)}\equiv \frac{1}{\sqrt{ N_-!}}\epsilon^{a_1\cdots a_{N_-}}\ket{u^-_{a_1}}\ket{u^-_{a_2}}\cdots\ket{u^-_{a_{N_-}}}\label{eq:fermi}
\ee
where the $\ket{u^-_{a}(\vec k)}$ form a basis of the $N_-$ filled bands (with negative energy). 
The metric for this state has also been considered previously in Ref.\onlinecite{Matsuura10}. We can interpret this state as the Fermi sea for fixed $\vec k$. Notice that when two or more of these negative energy states are degenerate (as will be the case in the Dirac models to be considered below), the basis is only defined modulo a $\vec k$ dependent "gauge rotation" $U(\vec k)\in U(N_-)$. Under this non-Abelian gauge transformation the state $\ket{\psi^{\rm val}}$ transforms with $\det U(\vec k)$ which is a pure phase, thus rendering the metric tensor completely invariant under such a basis redefinition.

 Physically, the quantum metric $g^{\rm val}_{\mu\nu}$ of the state $\ket{\psi^{\rm val}}$ defined by Eq.~(\ref{gab_multiple_lambda}) measures how much the unit vector $\ket{\psi^{\rm val}}$ has rotated in the $N_{-}$--particle Hilbert space as one moves from ${\bf k}$ to ${\bf k}+\delta{\bf k}$. 
It is possible to express $g^{\rm val}_{\mu\nu}$ explicitly in terms of one-particle states using standard techniques from second quantization.
The result is 
\be
g^{\rm val}_{\mu\nu}(\vec k)=
\frac{1}{2}\sum_a \biggl(\langle\partial_{\mu}u_{a}^-|Q_+|\partial_{\nu} u_a^-\rangle
+\langle\partial_{\nu}u_{a}^-|Q_+|\partial_{\mu}u_{a}^-\rangle\biggr)
\label{eq:gval}
\ee
where we have defined the projectors onto the positive and negative eigenstates: 
\bea
Q_{\pm}\equiv\sum_{a=1}^{N_\pm}\ket{u^\pm_a}\bra{u^\pm_a}\,, 
\label{eq:proj}
\eea
which satisfy $Q_++Q_-=1$ and $Q_\pm^2=Q_\pm$.
We futher define the spectrally flattened Hamiltonian $Q(\vec k)\equiv Q_+(\vec k)-Q_-(\vec k)$,
which has the same eigenstates as $H(\vec k)$ but with eigenvalues $\pm 1$.
The Hamiltonians $H(\vec k)$ and $Q(\vec k)$ can be continuously deformed into each other without closing the band gap, and thus have  identical topological properties. 
We remark that $ Q(\vec k)$ takes values in the complex Grassmannian $\frac{U(N)}{U(N_+)\times U(N_-)}$, since it can be specified by a diagonalizing unitary matrix $\in U(N)$ with two such matrices yielding the same $Q$ if they differ by a gauge transformation $\in U(N_+)\times U(N_-)$\cite{Ryu10}. This manifold has a canonical Riemannian metric, and  
 $g^{\rm val}(\vec k)$ is precisely the pullback of this metric to the BZ along the map $ Q(\vec k)$.
By differentiating
$Q_-\ket{u^-_a}=\ket{u^-_a}$ w.r.t.~$ k^\mu$, one obtains
$
Q_+\ket{\partial_\mu u^-_a}=\partial_\mu Q_-\ket{u^-_a}
$ 
and hence after some straightforward projector algebra
\be
g^{\rm val}_{\mu\nu}(\vec k)=\frac{1}{8}\tr\, \partial_\mu Q\, \partial_\nu Q\,.
\label{eq:gval1}
\ee
Indeed, $Q$ is manifestly invariant under the aforementioned basis redefinitions, and so is $g^{\rm val}$, hence $g^{\rm val}$ is in principle measurable. Detailed derivation of the above formalism is given in Appendix \ref{apx:detailed_formalism}.

 We will focus on the Dirac Hamiltonians that realize TIs and TSCs according to their symmetry classes,  which take the form
 \be
	H(\vec k)=\sum_{i=0}^D d^i(\vec k) \Gamma_i
\label{eq:dirac}
 \ee
 where ${\bf d}=(d^{0},d^{1}...d^{D})$ is the vector that parametrizes the Dirac Hamiltonian, and $\Gamma_i$ are  $N=2^n$ dimensional Dirac matrices satisfying the Clifford algebra $\{\Gamma_i,\Gamma_j\}=2\delta_{ij}$. 
The Hamiltonian in Eq.~(\ref{eq:dirac}) has $N/2=2^{n-1}$ degenerate eigenvalues $|\vec d|$ and $2^{n-1}$ eigenvalues $- |\vec d|$ (in our context corresponding to empty and filled bands respectively). 
The spectrally flattened Hamiltonian can be expressed as
\be
Q(\vec k)=\sum_{i=0}^D n^i(\vec k)\Gamma_i 
\label{Qk_spectral_flattened}
\ee
where  throughout the article we define the unit vector $\vec n(\vec k)\equiv \vec d(\vec k)/|\vec d(\vec k)|$. 



%

The image of the map $\vec n(\vec k)$ lies in the $D$ dimensional sphere $S^D\subset \mathbb R^{D+1}$,
 that we will refer to as the  {\em Dirac sphere} (DS) in what follows.
In Ref.~\onlinecite{vonGersdorff21_unification} it was shown that all nontrivial topological invariants can be expressed in terms of the degree  $\deg[\vec n]$ which was also referred to as the wrapping number. The latter counts how many times the BZ torus  wraps around the DS  under the map $\vec n(\vec k)$. In particular,  all known integer topological invariants are given by $\deg[\vec n]$ or $2\deg[\vec n]$, while all the binary ones are given by $(-1)^{\deg[\vec n]}$. One explicit representation of the wrapping number is given by \cite{vonGersdorff21_unification}
\be
\deg[\vec n]= \frac{1}{V_D}\int_{\rm BZ}J_{\vec n}(k)\,d^{D}k\,,
\label{eq:deg}
\ee
 where $V_D=2\pi^{\frac{D+1}{2}}/\Gamma(\frac{D+1}{2})$ is the volume of $S^D$, and the "curvature function"\cite{Chen16,Chen16_2,Chen17,Chen19} 
$J_{\bf n}$ is defined as
\be
J_{\vec n}(\vec k)\equiv\det \left(\vec n,\frac{\partial \vec n}{\partial k^1},\dots,\frac{\partial \vec n}{\partial k^D}\right)\equiv \det E_{\vec n},
\ee
which is sometimes referred to as the cyclic derivative of the $\vec n$-vector.



We proceed to relate the curvature function $J_{\vec n}$ to the quantum metric  $g_{\mu\nu}^{\rm val}$.
First we apply our general equation for the quantum metric of the valence band, Eq.~(\ref{eq:gval1}) to Dirac models to find
\be
g_{\mu\nu}^{\rm val}=
\frac{N}{8}\,\partial_\mu\vec n\cdot\partial_\nu \vec n
\label{eq:gvalDirac}
\ee
where we 
used Eq.~(\ref{Qk_spectral_flattened}) and applied the Clifford algebra.
This equation is remarkably simple. 
In fact, it is nothing but the pullback of the canonical metric of the $D$-sphere of radius $\sqrt{N/8}$ to the BZ along the map $\vec n(\vec k)$.
Next we write
\be
J_{\vec n}^2=\det E^T_{\vec n}E_{\vec n}
=\det\begin{pmatrix}
\vec n\cdot\vec n&\vec n\cdot \partial_\nu\vec n\\
\partial_\mu\vec n\cdot\vec n&\partial_\mu\vec n\cdot\partial_\nu\vec n
\end{pmatrix}
\\
=\det \partial_\mu\vec n\cdot\partial_\nu\vec n\,.
\label{eq:mqq0}
\ee
where it was used that $\vec n^2=1$ and $\vec n\cdot\partial_\mu\vec n=0$.
Combining Eq.~(\ref{eq:mqq0}) and Eq.~(\ref{eq:gvalDirac}) we arrive at one of our main results
\be
|J_\vec n|=\left(\frac{8}{N}\right)^\frac{D}{2}\sqrt{\det g^{\rm val}}
\label{eq:mcc}
\ee
a relation that we refer to as the metric-curvature correspondence. A similar result that relates the integrand of Chern number and winding number, individually, to the canonical metric on the Dirac sphere given by Eq.~(\ref{eq:gvalDirac}) has also been proposed recently\cite{Mera21}. Nevertheless, besides giving a more unified formalism, we emphasize that it is the link to the many-body quantum metric $g_{\mu\nu}^{\rm val}$ we uncovered that makes this correspondence experimentally relevant, as demonstrated below.

\subsection{Experimental implications} 

The metric-curvature correspondence,  Eq.~(\ref{eq:mcc}), prompts us to seek for a measurement for the momentum-profile of the quantum metric $ g_{\mu\nu}^{\rm val}({\bf k})$, since it may give direct information about the topological invariant in  any dimension and symmetry class. Out of several existing proposals \cite{Neupert13,Kolodrubetz13,Tan19,Gianfrate20}, we focus on the one based on time-dependent perturbation theory \cite{Ozawa18}, which has been verified experimentally in a single atom Rabi oscillation in NV centers in diamonds \cite{Chen20_Cappellaro,Yu19}. Our aim is to generalize this theory to our Dirac Hamiltonian that has $N/2$-fold degeneracy in both the filled and empty bands, and consider the application of a pulse electric field of magnitude ${\bf E}^{0}$ 
and pulse profile $g(t)$,
\be
\vec E(t)=\vec E^0 g(t)
\label{oscillating_E_field}
\ee
to the Bloch Hamiltonian $H(\vec k)$, and the pulse profile is assumed to satisfy $g(\pm\infty)=0$ but is  otherwise arbitrary. 
Electromagnetic gauge invariance
\be
\phi\to\phi-\partial_t\Lambda\qquad \vec A\to\vec A+\nabla\Lambda
\ee
allows for various equivalent implementations of the electric field, Eq.~(\ref{oscillating_E_field}). 
We find it convenient to work in the gauge 
\be
\vec A=0\qquad \phi = -g(t)\vec E_0\cdot \vec x\,.
\label{eq:gauge1}
\ee
We can then straightforwardly apply first-order time dependent perturbation theory for large times, yielding the transition amplitude  from the initial to the final state
\be
a_{i\to f}=\frac{ie}{\hbar}
\int_{-\infty}^\infty e^{i\omega t}g(t)\,
\braket{\phi_f|\vec E_0\cdot \vec x |\phi_i}
\ee
where $\hbar\omega\equiv \eps_f-\eps_i$. 
Suppose that an electron, initially in a filled-band Bloch state $\ket {u^-_a(\vec k)}$, under the influence of the electric field makes a transition to the empty-band state  $\ket {u^+_b(\vec k')}$. 
\footnote{Notice that we do not need to consider transitions within the fully occupied degenerate valence bands since they are forbidden by the Pauli principle at zero temperature. At finite temperature and for very small band gaps, a non-negligible number of valence electrons might be thermally excited and one would have to also consider intra-band transitions, which should be treated elsewhere.}
The dipole operator matrix elements between two general Bloch states is given by\cite{Karplus54}
\bea
&&\braket{u^{\sigma'}_b(\vec k')|\vec x|u^\sigma_a(\vec k)}\\
&&=-i\delta_{ab}\delta_{\sigma\sigma'}\nabla_{\vec k}\delta_{\vec k,\vec k'}
+i
\braket{u^{\sigma'}_b(\vec k)|\nabla_{\vec k} |u^\sigma_a(\vec k)}
\delta_{\vec k,\vec k'}\nn
\eea
and hence for $\sigma'=+$ and $\sigma=-$ we have simply
\footnote{More rigorously, we should consider the transition from the $n_{\rm val}$-particle valence band state $\ket{\phi_i}=\ket{\rm val}$, where $n_{\rm val}$ is the total number of valence electrons, to the state $\ket{\phi_f}=\ket{ \rm val'}=c_{\vec k',b}^{+\,\dagger} c_{\vec k,a}^-\ket{\rm val}$.
However, it is easily seen that 
$\braket{\rm val'| \vec x | \rm val}$ is precisely equal to the left hand side of
 Eq.(\ref{eq:mel}).}
\be
\braket{u^{+}_b(\vec k')|\vec x|u^-_a(\vec k)}=
i\braket{u^{+}_b(\vec k)|\nabla_{\vec k} |u^-_a(\vec k)}
\delta_{\vec k,\vec k'}
\label{eq:mel}
\ee
which is momentum conserving. 

Assuming that the pulse is only on for a finite amount of time, the first order perturbation theory yields the probability for such a transition:
\be
p_b^{(a)}(\vec k)=\left(\frac{e }{\hbar}\right)^2\,
\left|\tilde g(\omega(\vec k))\right|^2\,
 \left|\braket{u_b^+(\vec k)|\vec E^0\cdot\nabla_{\vec k}|u_a^-(\vec k)}\right|^2 
\label{nfkOmegat_delta}
\ee
where $\hbar \omega\equiv \eps_b-\eps_a$ and $\tilde g(\omega)\equiv\int_{-\infty}^\infty e^{i\omega t}g(t)dt$ denotes the Fourier transform of the pulse. We would like to briefly remark on the shape of the pulse $g(t)$: If it has a dominant frequency such that $\tilde g(\omega)$ is strongly peaked at some $\Omega$, only part of the filled band will be excited (those with $\omega\sim \Omega$)  and one might have to scan over different values of $\Omega$ to eventually cover the whole BZ. On the other hand, one could also envisage an extremely short pulse which has a rather flat profile $\tilde g(\omega)$ and thus the whole filled band will be democratically excited.


The transition probability into any of the conduction band states is then simply $p^{(a)}(\vec k)=\sum_{b\in c}p_{b}^{(a)}({\bf k})$. Suppose we direct the electric field to be along $\mu$ direction ${\bf E}^{0}=E^{0}{\hat{\boldsymbol\mu}}$, one finds
\begin{eqnarray}
p^{(a)}(\vec k)
=\left(\frac{e E^0}{\hbar}\right)^2\,
\left|\tilde g(\omega(\vec k))\right|^2\,
\sum_{b} \left|\braket{u_b^+(\vec k)|\partial_\mu u_a^-(\vec k)}\right|^2 \nn\\
\label{Gammak_gmunu}
\end{eqnarray}
For { the nondegenerate case} of a single valence band, this is directly proportional to the diagonal element of the quantum metric $g_{\mu\mu}^{\rm val}({\bf k})=g^{\psi}_{\mu\mu}({\bf k})$  of the single-particle Bloch state $|\psi\rangle=|u_1^-\rangle$\cite{Ozawa18}. However, for the degenerate case, the right hand side is no longer proportional to the quantum metric of any particular state, and moreover the above formalism is not gauge invariant. Instead, the truly gauge-invariant and measurable object is the probability summed over the  degenerate occupied valence bands 
\begin{eqnarray}
\nu({\bf k})&=&
\left(\frac{e E^0}{\hbar}\right)^2\,
\left|\tilde g(\omega(\vec k))\right|^2
\sum_{b\in c,a\in v} 
\left|\braket{u_b^+(\vec k)|\partial_\mu u_a^-(\vec k)}\right|^2 \nn\\
&=&\left(\frac{e E^0}{\hbar}\right)^2\,
\left|\tilde g(\omega(\vec k))\right|^2\,
 g_{\mu\mu}^{\rm val}({\bf k}),
\label{TrGamma_inv}
\end{eqnarray}
where we have used Eq.~(\ref{eq:gval}).
Similarly, we can consider two different driving protocols $E_{\mu}=E^{0}g(t)$ and $E_{\nu}=\pm E^{0}g(t)$ in different spatial directions. By subtracting the results of the two driving protocols
$\nu^{+}({\bf k})-\nu^{-}({\bf k})=
\left(\frac{e E_0}{\hbar}\right)^2\,
\left|\tilde g(\omega(\vec k))\right|^2\,
4
g_{\mu\nu}^{\rm val}({\bf k})$, one can obtain the off-diagonal elements of the metric\cite{Ozawa18}.


\begin{figure}[ht]
\begin{center}
\includegraphics[clip=true,width=0.9\columnwidth]{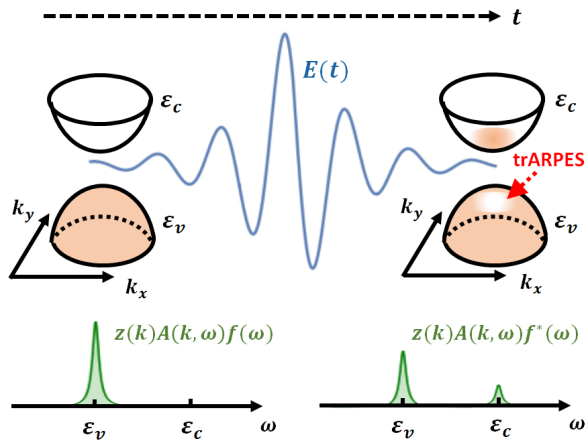}
\caption{ Schematics of the proposed trARPES experiment. All the particles are in the degenerate filled bands $\varepsilon_{v}$ before the pulse electric field ${\bf E}(t)$ is applied. Immediately after the pulse, some particles are in the degenerate empty bands $\varepsilon_{c}$, causing a violation of spectral sum rule in the filled bands at momentum ${\bf k}$, which can be measured by detecting the spectral function times distribution function $z({\bf k})A({\bf k},\omega)f^{\ast}(\omega)$ using trARPES, where the matrix element effect is taken care of by $z({\bf k})$. The quantum metric $g_{\mu\nu}({\bf k})$ can be extracted from the measurement according to Eqs.~(\ref{TrGamma_inv}) and (\ref{Gammaint_Akw_g}). } 
\label{fig:trARPES_AkEwt}
\end{center}
\end{figure}

 Encouraged by the result in Eq.~(\ref{TrGamma_inv}), we further propose the following protocol of trARPES experiment to measure the 
 quantity $\nu(\vec k)$.
 Our proposal is similar to measuring the depletion rate of the ultracold atom version of Chern insulator under periodic perturbation  \cite{Tran17,Asteria19}. Because in the summation of Eq.~(\ref{Gammak_gmunu}), only the final states not equal to the initial state $b\neq a$ contribute, this implies one should measure how many electrons are leaving the degenerate filled band states at momentum ${\bf k}$. In practice, how many electrons are leaving the  filled bands can be measured by the ARPES spectral function times the Fermi distribution, which now has been modified due to the pump pulse. 
 To illustrate this, consider a static Dirac model at zero temperature, which gives the imaginary time Green's function $G({\bf k},i\omega)^{-1}=i\omega-H({\bf k})$. After an analytical continuation $i\omega\rightarrow\omega+i\eta$ and taking the scattering rate $\eta\rightarrow 0$, one obtains the retarded Green's function $G^{ret}({\bf k},\omega)$ and subsequently the ARPES spectral function 
\begin{eqnarray}
&&A({\bf k},\omega)=-\frac{1}{\pi}{\rm Im}\left\{{\rm tr}\left[G^{ret}({\bf k},\omega)\right]\right\}
\nonumber \\
&&=\frac{N}{2}\delta(\omega-d({\bf k}))+\frac{N}{2}\delta(\omega+d({\bf k})),
\label{Akw_delta_fn}
\end{eqnarray}
signifying the $N/2$-fold degeneracy. Ideally, the spectral function satisfies the spectral sum rule $\int_{-\infty}^{\infty}d\omega\,A({\bf k},\omega)f(\omega)=N/2$, where $f(\omega)=\theta(-\omega)$ is the Fermi distribution at zero temperature. However, the ARPES experiment in reality contains the matrix element effect that affects the absolute scale of $A({\bf k},\omega)$ and renders this normalization ${\bf k}$-dependent. For our purpose of extracting quantum metric without ambiguity, one may add a ${\bf k}$-dependent factor $z({\bf k})$ to fit the spectral sum rule until it is rigorously satisfied at equilibrium at any momentum $z({\bf k})\int_{-\infty}^{\infty}d\omega\,A({\bf k},\omega)f(\omega)=N/2$, such that the matrix element effect does not obscure our formalism below. 



When the time dependent electric field ${\bf E}^{0}g(t)$ is applied, the spectral sum rule becomes time-dependent. The average number of electrons in the valence band, shortly after the pulse has ceased (such that relaxation has not yet set in) is
\begin{eqnarray}
&&z({\bf k})\int_{-\infty}^{0}d\omega\,A({\bf k},\omega)f^*(\omega)
=\frac{N}{2}-\sum_{a\in v}\sum_{b\in c}p^{(a)}_{b}({\bf k}),
\label{intAkwf_nf}
\end{eqnarray}
where we assume that the spectral function $z({\bf k})A({\bf k},\omega)$ remains a $\delta$-function, but the distribution function is no longer the equilibrium Fermi function $f^*(\omega)\neq\theta(-\omega)$ because the electron has a finite probability to enter other final states $b$. 
Then
\begin{eqnarray}
&&\nu(\vec k)
=\frac{N}{2}-z({\bf k})\int_{-\infty}^{0}d\omega\,A({\bf k},\omega)f^*(\omega).
\label{Gammaint_Akw_g}
\end{eqnarray}
Through comparing Eqs.~(\ref{TrGamma_inv}) and (\ref{Gammaint_Akw_g}), we see that the quantum metric can be extracted from trARPES by measuring the particle density loss in the degenerate filled bands at ${\bf k}$  immediately after the pulse, as shown schematically in Fig.~\ref{fig:trARPES_AkEwt}.
In Sec.\ref{sec:ARPES_graphene}, we use the measurement of topological charge of graphene to elaborate the feasibility of our proposal.  We also emphasize that our proposal based on Eqs.~(\ref{TrGamma_inv}) and (\ref{Gammaint_Akw_g}) is a universal protocol to measure the many-body quantum metric in any gapped degenerate fermionic systems, not only limited to topological materials. Moreover, ${\bf k}$ is not limited to momentum but can be any system parameters, which may also help to measure the fidelity susceptibility associated with quantum phase transitions in general\cite{You07,Zanardi07,Gu08,Yang08,Albuquerque10,Gu10,Carollo20}, provided the driving field ${\bf E}(t)$ couples to the system parameter  ${\bf k}$ in the same way as that described in Eqs.~(\ref{oscillating_E_field}) to (\ref{nfkOmegat_delta}). It is only through the metric-curvature correspondence in Eq.~(\ref{eq:mcc}) that the measurement performed on TIs and TSCs would directly reveal the topological order.

We now comment on several issues one may encounter in realistic ARPES measurements. Firstly, our proposal only allows to measure the modulus of the integrand $J_{\bf n}$ of the wrapping number ${\rm deg}[{\bf n}]$ via Eq.~(\ref{eq:mcc}), but not the sign of $J_{\bf n}$. In reality, the BZ consists of domains of different signs of $J_{\bf n}$, and to fix these signs to unambiguously determine the wrapping number requires some other input, such as band structure calculations. Nevertheless, the modulus of $J_{\bf n}$ itself already yields various valuable information about the topological order, such as the correlation length, scaling laws\cite{Chen17,Chen19_AMS_review}, and fidelity susceptibility\cite{Panahiyan20_fidelity_susceptibility,Molignini21_Kitaev_cross_dim}. Secondly, for systems beyond the Dirac model, it remains to be clarified how the metric-curvature correspondence will be modified, which may need to be dealt with case by case. Thirdly, many-body effects can broaden the spectral function and invalidate the sharp $\delta$-function in Eq.~(\ref{Akw_delta_fn}). Effects of this kind require a fully nonequilibrium many-body version of our formalism, which awaits further investigations. Finally, since ARPES is a surface probe, our proposal is presumably more suitable to detect the topology of 2D systems, provided the ARPES laser spot is smaller than the systems size such that the edge states do not interrupt. For 3D systems, most likely the method will be detecting the topology of the surface states if the system is in the topologically nontrivial phase.

\subsection{ARPES protocol applied to graphene \label{sec:ARPES_graphene}}

\begin{figure}[ht]
\begin{center}
\includegraphics[clip=true,width=0.99\columnwidth]{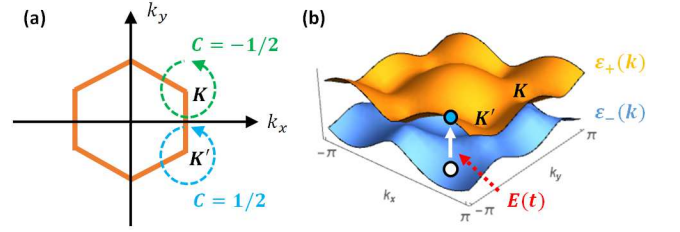}
\caption{(a) The topological charge ${\cal C}=\pm 1/2$ obtained by integrating the Berry connection along a circle of arbitrary radius circulating the Dirac points ${\bf K}$ and ${\bf K}'$. The orange line denotes the BZ. The corresponding quantum metric in the angular direction should be a constant $g_{\phi\phi}^{\rm val}=1/2$. (b) Schematics of using a pulse electric field to excite the electrons from valence to conduction band. } 
\label{fig:graphene_trARPES}
\end{center}
\end{figure}

As a concrete example, we discuss the aforementioned ARPES technique applied to measuring the topological charge of graphene,  which is a topological semimetal that contains two Dirac points ${\bf K}$ and ${\bf K}'$ that have opposite topological charges.  For a $D$-dimensional topological semimetal in general, our wrapping number formalism still applies, but the integration in Eq.~(8) is over a compact $(D-1)$-dimensional surface enclosing each nodal point, and the metric-curvature correspondence is defined on this $(D-1)$-dimensional surface, as we shall see below for graphene with $D=2$. We choose graphene because its spectral function is extremely sharp due to the long mean free time  \cite{Bostwick07,Sprinkle09}, and moreover the spin degeneracy is well preserved since spin-orbit coupling is negligible\cite{HuertasHernando06,Min06,Yao07}. Consider only one spin species and expand the Hamiltonian around the two Dirac points 
\begin{eqnarray}
{\bf K}=\left(\frac{2\pi}{3},\frac{2\pi}{3\sqrt{3}}\right)\;,\;\;\;{\bf K}'=\left(\frac{2\pi}{3},-\frac{2\pi}{3\sqrt{3}}\right)
\end{eqnarray}
yields the linear Dirac Hamiltonian \cite{Bernevig13}
\begin{eqnarray}
&&H_{0}({\bf K}+{\bf k})=\frac{3}{2}t\left(k_{y}\sigma_{x}-k_{x}\sigma_{y}\right),
\nonumber \\
&&H_{0}({\bf K}^{\prime}+{\bf k})=\frac{3}{2}t\left(-k_{y}\sigma_{x}-k_{x}\sigma_{y}\right)\;,
\end{eqnarray}
where $t$ is the nearest-neighbor hopping on the honeycomb lattice. This linear dispersion is well satisfied up to energy $\sim 1$eV away from the Dirac point, which covers a large enough momentum space to perform the proposed pump-probe experiment\cite{Gierz13}. The eigenenergies and eigenstates  for one spin species, say spin up, are
\begin{eqnarray}
&&\varepsilon_{\pm}^{\bf K}({\bf k})=\pm\frac{3}{2}t_{1}k,\;\;\;|u_{\bf k\pm}^{\bf K}\rangle=\frac{1}{\sqrt{2}}\left(
\begin{array}{c}
1 \\
\mp ie^{i\phi}
\end{array}
\right),
\nonumber \\
&&\varepsilon_{\pm}^{\bf K'}({\bf k})=\pm\frac{3}{2}t_{1}k,\;\;\;|u_{\bf k\pm}^{\bf K'}\rangle=\frac{1}{\sqrt{2}}\left(
\begin{array}{c}
1 \\
\mp ie^{-i\phi}
\end{array}
\right),
\end{eqnarray}
 where $\phi$ is the polar angle of the momentum ${\bf k}=(k,\phi)$.
Integrating the valence band Berry connection $\langle u_{\bf k-}^{\bf K}|i\partial_{\varphi}|u_{\bf k-}^{\bf K}\rangle$ along a loop of radius $k$ circulating the Dirac points yields the topological charges
\begin{eqnarray}
&&\frac{1}{2\pi}\oint d\phi\langle u_{\bf k-}^{\bf K}|i\partial_{\phi}|u_{\bf k-}^{\bf K}\rangle=-\frac{1}{2\pi}\oint d\phi\langle u_{\bf k-}^{\bf K'}|i\partial_{\phi}|u_{\bf k-}^{\bf K'}\rangle
\nonumber \\
&&=-1/2,
\label{eq:grapheneC}
\end{eqnarray}
as shown schematically in Fig.~\ref{fig:graphene_trARPES} (a). The two spin degrees of freedom are completely decoupled and the Hamiltonian is block-diagonal. In our choice of  basis this means that the valence band metric reduces to the sum of the metrics of the spin-up and spin-down states $g_{\phi\phi}^{\rm val}=g_{\phi\phi}^{\uparrow}({\bf k})+g_{\phi\phi}^{\downarrow}({\bf k})$.
Focusing on the ${\bf K}$ point $|u\rangle\equiv|u_{\bf k-}^{\bf K}\rangle$, the Dirac Hamiltonian $H({\bf K+k})=d_{1}\sigma_{1}+d_{2}\sigma_{2}$ gives a quantum metric
\begin{eqnarray}
&&g^u_{\phi\phi}=\langle\partial_{\phi}u|\partial_{\phi}u\rangle-\langle\partial_{\phi}u|u\rangle\langle u|\partial_{\phi}u\rangle
=|\langle u|i\partial_{\phi}|u\rangle|^{2}
\nonumber \\
&&=\left|\frac{1}{2}\epsilon^{ab} {\bf n}_{a}\partial_{\phi}{\bf n}_{b}\right|^{2}=\frac{1}{4}.
\end{eqnarray}
We see that indeed the metric-curvature correspondence  is satisfied with $\sqrt{g^{\rm val}}=\sqrt{g^{\rm  val}_{\phi\phi}}=1/\sqrt{2}$. Therefore, if the proposed ARPES experiment yields a constant quantum metric in the angular direction $g^{\rm val}_{\phi\phi}=1/2$ at any momentum ${\bf k}$, then the topological charge is verified. 

Although it may be difficult to directly measure $g_{\phi\phi}$, one may apply an oscillating electric field in planar directions to extract $\left\{g_{xx},g_{xy},g_{yy}\right\}$, and then use the conversion between the derivatives $\partial_{k}=\cos\phi\partial_{x}+\sin\phi\partial_{y}$ and $\partial_{\phi}=-k\sin\phi\partial_{x}+k\cos\phi\partial_{y}$
to extract $g_{\phi\phi}$ from $\left\{g_{xx},g_{xy},g_{yy}\right\}$ by
\begin{eqnarray}
g_{\phi\phi}=k^{2}\sin^{2}\phi\,g_{xx}-k^{2}\sin 2\phi\,g_{xy}+k^{2}\cos^{2}\phi\,g_{yy}.\;\;\;\;\;
\label{gphiphi_to_gxy}
\end{eqnarray}
 To give an order of magnitude estimation, we use the numbers in the pump-probe experiment performed on graphene\cite{Gierz13}.  To measure the quantum metric, we suppose that the pump pulse excites the valence electron at ${\bf k}$ that has a band gap of the order of an electron volt $\hbar\omega\sim\varepsilon_{+}-\varepsilon_{-}\sim\,$eV, which corresponds to a frequency $\sim 10^{15}\,$Hz, as shown schematically in Fig.~\ref{fig:graphene_trARPES} (b).
For the matrix element we can estimate
\be
\braket{u_+|\partial_\mu u_-}=
\Braket{u_+|\frac{\partial_\mu H}{\eps_+-\eps_-}|u_-}\sim \frac{v_F}{\omega}
\ee
where $v_{F}\sim 10^{6}$m/s is the Fermi velocity of graphene. If the pulse is on for a short time $T\sim 0.1\, $ps, then $\tilde g(\omega)\sim T$ and 
\be
\nu(\vec k)\sim \left(\frac{eE^0v_F T}{\hbar\omega}\right)^2
\ee
Suppose we aim to excite $\nu(\vec k)\sim 10\%$ electrons after the pulse. This requires an electric field square of the order of $(E^{0})^{2}\sim 10^{13}$V$^{2}/$m$^{2}$. The fluence after applying the pulse is
\begin{eqnarray}
F\sim \frac{c\,\varepsilon_{0}}{2}|E^{0}|^{2}T\sim 10^{-4}\frac{\rm mJ}{\rm cm^{2}},
\end{eqnarray} 
where $c$ is the speed of light and $\varepsilon_{0}$ is the vacuum permittivity. This fluence is much smaller than that delivered in the graphene pump-probe experiment $\sim{\rm mJ}/{\rm cm^{2}}$\cite{Gierz13}, hence should be easily achievable and heating can be ignored. In fact, the pump-probe trARPES experiment has already revealed a significant amount of excitation from the valence to the conduction band, although the pump pulse ${\bf E}(t)$ is usually not polarized. Finally, we remark that the light-matter interactions\cite{Schuler21} and light-induced anomalous Hall effect\cite{Sato19} in graphene, as well as the nonequilibrium response of graphene in the pump-probe setup in the presence of many-body interactions has been addressed previously\cite{Schuler20}. Precisely how these realistic factors, in particular the many-body interactions, can affect the measurement of quantum metric is a fundamental issue that requires further investigations.

\section{Conclusions} 

In summary, we elaborate the metric-curvature correspondence between the modulus of the integrand of the wrapping number and the quantum metric of an appropriate many-body Bloch state described by Eq.~(\ref{eq:mcc}). Based on the validity of this correspondence in any dimension and symmetry class, we propose an ARPES measurement protocol to ubiquitously detect the momentum profile of the quantum metric, from which information about the topological properties of the system can be extracted. By generalizing a recently proposed time-dependent perturbation theory to degenerate bands, our proposal suggests to measure the violation of spectral sum rule caused by a pulse electric field to extract the quantum metric. Various complications in reality, such as systems beyond Dirac models or containing electronic correlations, requires further generalization of our formalism that awaits to be explored.

\acknowledgements 

The authors acknowledge stimulating discussions with M. Manske, S. Panahiyan, L. Kemper, B. Mera, and the experimental input from S. Kaiser. W. C. is financially supported by the productivity in research fellowship from CNPq.

\appendix

\section{Detailed derivation of the quantum metric formalism \label{apx:detailed_formalism}}

We now give some detailed calculation for the quantum metric. We would like to compute the metric for the $N_-$ particle state $|\psi^{\rm val}(\vec k)\rangle$ defined in Eq.~(\ref{eq:fermi}), i.e.
\begin{eqnarray}
g^{{\rm val}}_{\mu\nu}({\bf k})&=&\frac{1}{2}\langle\partial_{\mu}\psi^{\rm val}|\partial_{\nu}\psi^{\rm val}\rangle
+\frac{1}{2}\langle\partial_{\nu}\psi^{\rm val}|\partial_{\mu}\psi^{\rm val}\rangle
\nonumber \\
&&-\langle\partial_{\mu}\psi^{\rm val}|\psi^{\rm val}\rangle\langle\psi^{\rm val}|\partial_{\nu}\psi^{\rm val}\rangle.
\label{eq:A0}
\end{eqnarray} 
The simplest approach is to use second-quantization formalism with fermionic annihilation operators $c_{a,\sigma}$ where $\sigma=\pm$ denotes positive and negative energy eigenstates respectively.
Then by the standard formula for multi-particle operators we have
\begin{eqnarray}
\frac{\partial}{\partial k^\mu} = \sum_{\substack{\sigma=\pm\\ \sigma'=\pm}}\sum_{a=1}^{N_{\sigma}}\sum_{a'=1}^{N_{\sigma'}}
\langle u_{a'}^{\sigma'}|\partial_\mu|u_{a}^{\sigma}\rangle
c_{a'\sigma'}^\dagger c_{a\sigma}.
\end{eqnarray} 
Acting with this on $|\psi^{\rm val}\rangle
=\prod_{a=1}^{N_-}c^\dagger_{a-}|0\rangle$, one gets
\begin{eqnarray}
|\partial_\mu\psi^{\rm val}\rangle&=&\sum_{a=1}^{N_-}\biggl(\langle u_{a}^-|\partial_\mu u_{a}^-\rangle+\sum_{a'=1}^{N_+}\langle u_{a'}^+|\partial_\mu u_{a}^-\rangle c_{a'+}^\dagger c_{a-}
\biggr)|\psi^{\rm val}\rangle.
\nonumber \\
\label{eq:A1}
\end{eqnarray} 
Then it is straightforward to compute
\begin{eqnarray}
\langle\psi^{\rm val}|\partial_\mu\psi^{\rm val}\rangle&=&\sum_{a=1}^{N_-}\langle u_{a}^-|\partial_\mu u_{a}^-\rangle
\nonumber \\
\langle\partial_\mu \psi^{\rm val}|\partial_\nu\psi^{\rm val}\rangle&=&
\left(\sum_{a=1}^{N_-}\langle\partial_\mu u_{a}^-| u_{a}^-\rangle\right)
\left(\sum_{a=1}^{N_-}\langle u_{a}^-|\partial_\nu u_{a}^-\rangle\right)
\nonumber \\
&&+\sum_{a=1}^{N_-}\langle\partial_\mu u_{a}^-|Q_+|\partial_\nu u_{a}^-\rangle,\;\;\;\;\;\;
\label{eq:A2}
\end{eqnarray} 
where $Q_+$ was defined in Eq.~(\ref{eq:proj}).
Applying Eqs.~(\ref{eq:A1}) and (\ref{eq:A2}) to Eq.~(\ref{eq:A0}), we obtain Eq.~(\ref{eq:gval}).

To further express $g_{\mu\nu}^{\rm val}$ in terms of the spectrally flattened Hamiltonian $Q({\bf k})$, we use $Q_{+}^{2}=Q_{+}$, $Q_{\pm}=(1\pm Q)/2$, and $\left(\partial_{\mu}Q_{-}\right)|u_{a}^{-}\rangle=Q_{+}|\partial_{\mu}u_{a}^{-}\rangle$ to write
\begin{eqnarray}
g_{\mu\nu}^{\rm val}&=&\frac{1}{2}\sum_{a=1}^{N_{-}}\left(\langle u_{a}^{-}|\partial_{\mu}Q_{-}\partial_{\nu}Q_{-}|u_{a}^{-}\rangle+\langle u_{a}^{-}|\partial_{\nu}Q_{-}\partial_{\mu}Q_{-}|u_{a}^{-}\rangle\right)
\nonumber \\
&=&\frac{1}{2}{\rm tr}\left(Q_{-}\partial_{\mu}Q_{-}\partial_{\nu}Q_{-}+Q_{-}\partial_{\nu}Q_{-}\partial_{\mu}Q_{-}\right)
\nonumber \\
&=&\frac{1}{2}{\rm tr}\,\partial_{\mu}Q_{-}\partial_{\nu}Q_{-}^{2}
\nonumber \\
&=&\frac{1}{2}{\rm tr}\,\partial_{\mu}\left[\frac{1}{2}(1-Q)\right]\partial_{\nu}\left[\frac{1}{2}(1-Q)\right]
\nonumber \\
&=&\frac{1}{8}{\rm tr}\,\partial_{\mu}Q\,\partial_{\nu}Q,
\end{eqnarray}
which gives Eq.~(\ref{eq:gval1}).\cite{Matsuura10}

\bibliography{Literatur}

\end{document}